\documentclass[journal = jccp]{achemso}
\setkeys{acs}{usetitle = true}

\usepackage{amssymb,amsmath,bm}
\usepackage{graphics}
\usepackage{graphicx}
\usepackage[all]{xy}

\author{Marisa N. Faraggi,$^{1,2}$, Vitaly N. Golovach,$^{3,4,5}$ Sebastian Stepanow,$^{6,7}$, Tzu-Chun Tseng,$^{7}$  Nasiba Abdurakhmanova,$^{7}$  Christopher Seiji Kley,$^{7}$  Alexander Langner,$^{7}$  Violetta Sessi,$^{8}$ Klaus Kern$^{7,9}$  and Andres Arnau$^{1,3,4}$ }
\email{andres.arnau@ehu.es;  phone:  +34-943018204}
\affiliation{$^1$Donostia International Physics Center (DIPC), P.
de Manuel Lardizabal 4, E-20018 San Sebasti\'{a}n, Spain \\
$^2$ Instituto de Astronomia y F\'{\i}sica del Espacio, Conicet, Bs.As. Argentina\\
$^3$Departamento de F\'{\i}sica de Materiales, Facultad de Ciencias
Qu\'{\i}micas, Universidad del Pa\'{\i}s Vasco, Apdo. 1072,
E-20080 San Sebasti\'an, Spain \\
$^4$Centro de F\'{\i}sica de Materiales CFM, Materials Physics
Center MPC, Centro Mixto CSIC-UPV/EHU,
P. de Manuel Lardizabal 5, E-20018 San Sebasti\'{a}n, Spain \\
$^{5}$ IKERBASQUE, Basque Foundation for Science, E-48011, Spain\\ 
$^6$ Department of Materials, ETH Z\"urich, H\"onggerbergring 64, 8093 Z\"urich, Switzerland\\
$^7$ Max Planck Institute for Solid State Research, Heisenbergstrasse 1, 70569 Stuttgart, Germany\\
$^8$ European Synchrotron Radiation Facility, BP 220, 38043 Grenoble, France\\
$^{9}$ Institut de Physique de la Mati\`{e}re Condens\'{e}e, \'{E}cole Polytechnique F\'{e}d\'{e}rale de Lausanne, CH-1015 Lausanne, Switzerland\\
}

\title{Modelling Ferro- and Antiferromagnetic Interactions in Metal-Organic Coordination Networks }

\begin{document}
\begin{abstract}
 
Magnetization curves of two rectangular metal-organic coordination networks formed by the organic ligand TCNQ (7,7,8,8-tetracyanoquinodimethane) and two different (Mn and Ni) 3d transition metal atoms [M(3d)] show marked differences
that are explained using first principles density functional theory and model calculations. We find that the existence of a weakly dispersive hybrid band with M(3d) and TCNQ character crossing the Fermi level is determinant for the appearance of ferromagnetic coupling between metal centers, as it is the case of the metallic system Ni-TCNQ  but not of the insulating system Mn-TCNQ. The spin magnetic moment localized at the Ni atoms induces a significant spin polarization in the organic molecule; the corresponding spin density being delocalized along the whole system. The exchange interaction between localized spins at Ni centers and the itinerant spin density is ferromagnetic. Based on two different model Hamiltonians, we estimate the strength of exchange couplings between magnetic atoms for both Ni- and Mn-TCNQ networks that results in weak ferromagnetic  and very weak antiferromagnetic correlations for Ni- and Mn-TCNQ networks, respectively.

\end{abstract}

%
%

%
\section{\protect\bigskip Introduction}

Understanding the magnetic behavior of low dimensional systems is a challenge that has recently given rise to a number of works. \cite{gambardella2002, gambardella2003,bluegel2011, katsnelson2013}  Additionally, several studies have proposed systems showing high temperature ferromagnetism \cite{chen2008, xie2011,Schwingen2013, naveh2013,pantelides2013, flipse2013}.  However, in general, it is hard to predict the type, strength and range of magnetic interactions responsible for the existence of magnetic order.  The kind of systems that have been explored in recent years is rather vast,  ranging from substitutional magnetic impurities in graphene, \cite{santos2010} dilute magnetic semiconductor nanocrystals, \cite{ayuela}  hydrogenated epitaxial graphene \cite{xie2011} to molecular magnets. \cite{miller2011} In particular, bulk molecular crystals \cite{miller2012} are especially attractive to us because  two-dimensional  (2D) metal-organic  coordination networks (MOCN) on surfaces can be considered their analogues, as coordination chemistry compounds.

Of special interest is the growth of monolayer  films on single crystal surfaces using self-assembly techniques to form 2D coordination networks made of 3d transition metal atoms and organic ligands \cite{uta2007, umbach2012, giovanelli2014}. This permits to achieve a relatively high surface density of magnetic moments, localized at the 3d transition metal atom centers and forming a regular 2D structure with the organic ligands. In this way, metal atom cluster formation is avoided. However, critical temperatures in low dimensional systems are known to be much lower than in bulk three dimensional crystals \cite{sarma2005,miller2012}. Indeed,  2D isotropic systems with finite range exchange interaction cannot show long range ferromagnetic order at finite temperatures \cite{mermin_wagner,bruno2001}. 

In this work we study the low temperature magnetic behavior of MOCNs formed by self-assembly of 3d transition metal atoms and strong acceptor molecules on surfaces. In particular, we focus on the case of rectangular lattices with 1:1 stoichiometry and 4-fold coordination, that are known to form on metal surfaces, like Ag(100) or Au(111) \cite{nasiba2013}. Those structures represent easily accessible and tunable experimental realizations of electronic correlated systems and are, therefore, also interesting from a fundamental point of view. 

Previous studies \cite{faraggi2012,nasiba2013}  suggest that, in the case of non-reactive surfaces like Au(111), the underlaying substrate on top of which the metal-organic coordination network is grown plays only a minor role in determining the overlayer electronic properties, such as the type of bonding and coordination between  the 3d metal centers and the organic ligands. This is due to the formation of strong lateral bonds between the metal atoms and the organic molecules, which lift up the metal atoms from the surface and reduce, consequently,  the surface to metal interaction  \cite{nasiba2013,gambardella2009}. However, there are other metal surfaces, such as Cu(100), in which a significant charge transfer between the surface and the metal-organic network takes place \cite{cerda2010}. 

We specifically wonder whether this minor role of the substrate still holds for the magnetic interaction between the 3d transition metal atom spins when they are embedded in a 2D MOCN, including the sign, strength, and range of the spin-spin coupling, as compared to the case of 3d transition metal impurities on metals, where metal surface electrons mediate RKKY-type interactions \cite{yosida}. In principle, for the same organic ligand, stoichiometry and coordination, one could expect that the particular 3d transition metal atom center in the 2D MOCN is determinant in the type of  magnetic interaction (FM or AFM) depending on the 3d manifold energy level structure close to the Fermi level. As shown below, our results based on density functional theory (DFT) calculations at T=0 confirm that this is indeed the case because they permit to explain the observed trends in the measured X-ray magnetic circular dichroism (XMCD) data with the help of two model Hamiltonians. 

\bigskip

\section{\protect\bigskip Results and Discussion}

\begin{figure}[!htb]
  \centering
   \includegraphics*[width=.75\textwidth] {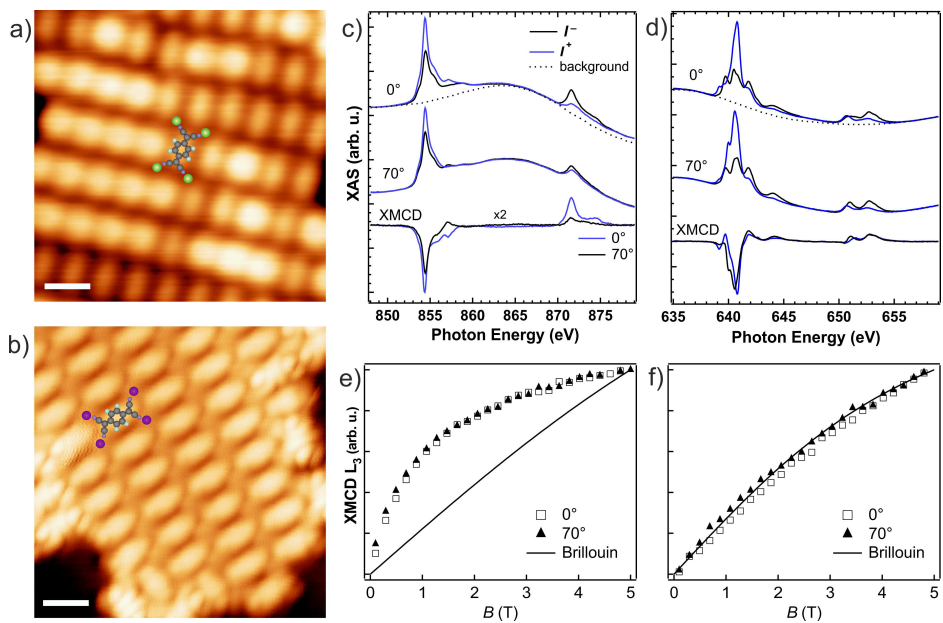}
    \caption{\label{FIG1}(color online)(a-b) STM images of (a) Ni-TCNQ and (b) Mn-TCNQ networks self-assembled on Au(111). The model for the unit cell structure is superposed to the images. (Scale bar in both images = 1 nm) (c-d) XAS and corresponding XMCD spectra for © Ni-TCNQ and (d) Mn-TCNQ networks for normal $(0^\circ)$ and grazing $(70^\circ)$ x-ray incidence angles. Note that because of the low coverage the metal L-edges are superposed to the XAS background of the substrate (shown for normal incidence). (T = 8 K, B = 5 T; XMCD: $0^\circ$=blue and $70^\circ$=black). (e-f) Magnetization curves for (e) Ni-TCNQ and (f) Mn-TCNQ obtained as the $L_3$ peak height vs magnetic field (T = 8 K) at normal (squares) and grazing incidence (solid triangles). For comparison the magnetization curves were normalized to 1 at B = 5 T.). The curves labeled Brillouin in (e) and (f) correspond to the paramagnetic behavior for S=1(e) and S=5/2 (f), respectively, at T=8 K (see the text). }
\end{figure}

Figures \ref{FIG1} a) and b) show STM topographical images of Ni- and Mn-TCNQ networks with a stoichiometry of 1$:$1 on Au(111), respectively. Each molecule forms four bonds to metal atoms via its cyano groups. Details of the structures can be found in Refs. \cite{faraggi2012,nasiba2013}. Figures \ref{FIG1}c) and d) show x-ray absorption (XAS) spectra recorded at the metal $L_{2,3}$-edge for parallel ($I^{+}$) and antiparallel ($I^{-}$) alignment of the photon helicity with the magnetic field B at normal ($\sim 0^\circ$) and grazing ($\sim 70^\circ$)  x-ray incidence. The corresponding XMCD spectra, defined as $I^{-}-I^{+}$, are shown at the bottom of the panels. Note, that because of the low coverage the data is superposed to a temperature dependent extended x-ray absorption fine structure background of the substrates. Background data is exemplarily shown for normal incidence. The metal coverage is estimated to 0.03 monolayers for the networks, one monolayer being one metal atom per site in the Au(111) top most layer. Both metal centers show pronounced fine structure of the white lines which originate from atomic multiplets of the final state configurations. This signifies electronic decoupling from the metal substrate and the formation of well-defined coordination bonds to the TCNQ molecules. The anisotropy in the XAS lineshape between normal and grazing incidence reflects the low symmetry environment of the metal centers. The XAS lineshapes of the Ni and Mn centers are compatible with $d^8$ and $d^5$ electronic configurations, respectively.\cite{nasiba2013,vdLaan1,vdLaan2} Thus, we expect unquenched spin moments of S=1 and S=5/2 for Ni and Mn, respectively, as evidenced also by the sizable XMCD intensity.

The possible magnetic interaction between the individual metal centers is revealed in the magnetization curves obtained as the XMCD $L_3$ peak \cite{L3} intensity (T = 8 K) normalized to 1 at B = 5 T for comparison (see Fig. \ref{FIG1}e,f). For both structures the magnetic susceptibility shows no strong apparent anisotropy. However, for the Ni-TCNQ network the curves show a stronger S-shape compared to Mn-TCNQ. This indicates ferromagnetic coupling between the Ni atoms, since we expect a smaller spin moment of S=1 for Ni compared to S=5/2 for Mn. Further insight can be drawn from the analysis of the shape of the magnetization curves by comparing them to the Brillouin function \cite{Brillouin} of the respective spin moment. The curves labeled Brillouin have been added to the panels \ref{FIG1}e) and f) with S=1 and S=5/2, respectively, assuming an isotropic g=2 factor. 
This approximation is based on the fact that in our systems the orbital moment is either isotropic (Ni) or very small (Mn). In neither case, can the g-factor account for the observed shape in the magnetization curves.
The Ni magnetization curves differ clearly from the paramagnetic S=1 susceptibility, whereas the Mn ions follow more closely the expected S=5/2 behavior. Our first principles and calculations and subsequent estimates of the exchange coupling constants using model Hamiltonians 
are consistent with this observations.

\begin{figure}[!htb]
  \centering
   \includegraphics*[width=.75\textwidth] {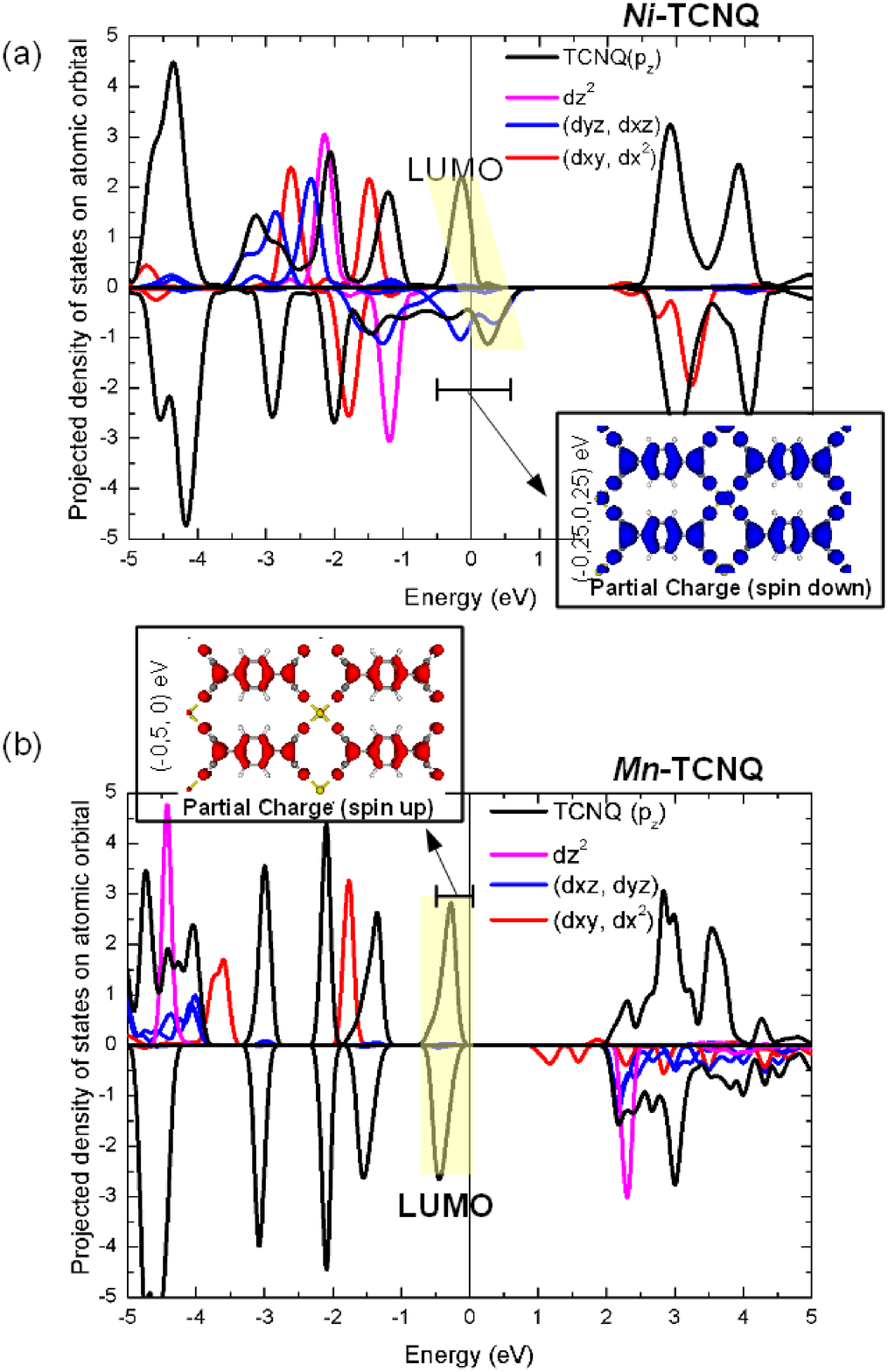}
    \caption{\label{FIG2a}(color online) Projected density of states [states/eV] onto metal atom centers ($3d$) [purple ($3d_{z^2}$), red ($3d_{xy}$~and~$3d_{x^2-y^2}$),  and blue lines ($3d_{xz}$~and~$3d_{yz}$) and TCNQ ($p_z$) [black line] orbitals for the (a) Ni-TCNQ and (b) Mn-TCNQ networks. The insets in pannels (a) and (b) show isocontours of constant electronic charge in a narrow energy range around the Fermi level  (partial charge), showing two different situations for Ni-TCNQ and Mn-TCNQ. The TCNQ LUMO is clearly seen in both cases, while only for Ni-TCNQ the minority spin $3d_{xz}$ orbital can be identified.}
\end{figure}

\begin{figure}[!htb]
  \centering
   \includegraphics*[width=.75\textwidth] {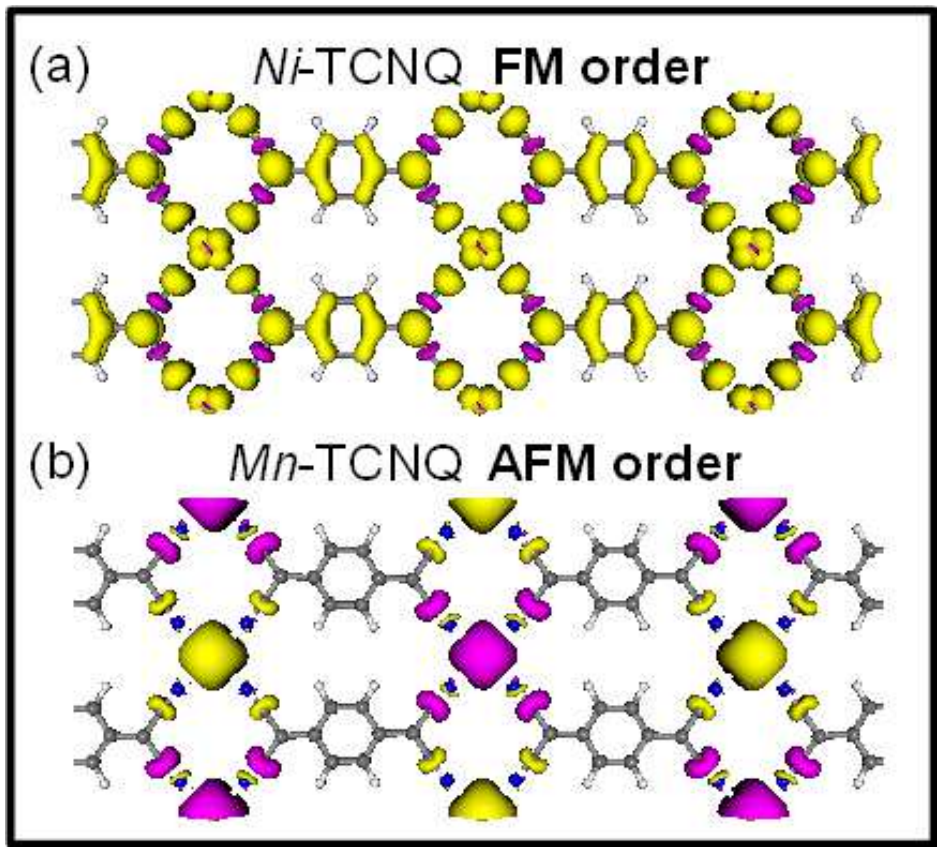}
    \caption{\label{FIG3a}(color online) Spatial distribution of the spin density in a rectangular checker-board 2x1 supercell of (a) Ni-TCNQ showing FM coupling between Ni atoms and spin polarization of the TCNQ LUMO and (b)Mn-TCNQ showing AFM coupling between Mn atoms and no spin polarization of the TCNQ LUMO.}
\end{figure}

Next we discuss the results from DFT calculations for both systems: Ni-TCNQ and Mn-TCNQ free standing overlayers excluding the Au(111) metal substrate. 
The free-standing-overlayer approximation, i.e., the neglect of Au(111) in our first principles calculations, is based on our previous finding \cite{faraggi2012}  of weak coupling between Mn-TCNQ overlayers to Au(111), whose direct  fingerprint is the observation of the herring bone reconstruction after the Mn-TCNQ network is grown on Au(111).
We focus first on the projected density of states (PDOS) onto different 3d metal atom orbitals, as well as onto TCNQ($p_z$) that permit to identify molecular orbitals close to the Fermi level, like the lowest unoccupied molecular orbital (LUMO). The 2D planar structure is located in the XY plane. Figure \ref{FIG2a} (a) shows the calculated PDOS for Ni-TCNQ. All the Ni($3d$) majority spin states are occupied, while one minority spin state remains completely empty [Ni($3d_{xy}$)]. Two other minority spin states  [Ni($3d_{xz}$) and Ni($3d_{yz}$)] are partially occupied and hybridize with the TCNQ LUMO, the corresponding dispersive band crosses the Fermi level [see Eq.~(\ref{eqspectrum})]. There is also a significant charge transfer from the Ni atom to the TCNQ LUMO of about one electron yielding a spin-polarized molecular state. As a consequence, there is a localized S=1/2 spin magnetic moment on the Ni atom and a somewhat smaller magnetic moment delocalized on the whole Ni and TCNQ system, as shown in Figure  \ref{FIG3a}. The inset in Figure  \ref{FIG2a} (a) illustrates the hybridization between the TCNQ LUMO and Ni($3d_{xz}$) orbitals. Therefore, for the Ni-TCNQ network our DFT calculations show that: (i) the system is metallic; it has a finite DOS at the Fermi level, (ii) there is a significant amount of hybridization between minority  Ni(3d) states and the TCNQ LUMO [a dispersive hybrid band crosses the Fermi level], and (iii) the TCNQ LUMO is spin polarized. This is a first hint for the existence of ferromagnetism in this system but it requires a further analysis (see next section Model for Ni-TCNQ ferromagnetism).

However, the situation is completely different in Mn-TCNQ. As shown in Figure \ref{FIG2a} (b),  all the Mn($3d$) majority spin states are occupied, while all the minority spin states remain empty, and none of them hybridize appreciable with the TCNQ LUMO [see the inset]. Additionally, the TCNQ LUMO is fully occupied due to a large electron transfer from the Mn atoms of practically two electrons and, therefore, the DOS at the Fermi level is negligible, i.e., the system is insulating. The spin magnetic moments are localized on the Mn atoms, as shown in Figure  \ref{FIG3a} (b),  and are very close to S=5/2. Therefore, the argument mentioned above as a hint for the existence of ferromagnetism in Ni-TCNQ does not apply for  Mn-TCNQ.
The reason for the different charge transfer to TCNQ LUMO from Mn and Ni metal centers, higher (and close to two electrons) in Mn-TCNQ than in Ni-TCNQ (about 1.3 electrons),  is that in Ni-TCNQ there is an important hybridization between the minority spin Ni($3d_{xz}$ and the TCNQ LUMO states, absent in the case of Mn-TCNQ.

Now we turn to the analysis of the coupling between the magnetic moments of the Ni and Mn atoms in their corresponding networks. We start by doing DFT calculations in a double size (2x1) supercell that contains two metal atoms in a checker-board configuration, so that we can treat both parallel (FM) and antiparallel (AFM) alignment of spins. For the Ni-TCNQ network we find that the FM configuration is energetically favored by 105.7 meV, while for the Mn-TCNQ network the AFM configuration is more favorable by 8.75 meV, per surface unit cell (2x1). Taking into account that the Mn atoms spin magnetic moment is five  times larger than that of the Ni atoms, we see that the coupling in the Mn-TCNQ system is two orders of magnitude smaller, and of opposite sign, as compared with Ni-TCNQ.  The corresponding spin densities are shown in Figure \ref{FIG3a} (a) and \ref{FIG3a} (b) for Ni-TCNQ and Mn-TCNQ, respectively, exhibiting rather different behavior. The spin density is delocalized all along the Ni atoms and TCNQ molecule (a) , while it is localized at the Mn atoms sites (b). To understand the correlation between magnetic coupling and chemical bonding in the two systems, next we describe two models that explain the mechanism for ferromagnetism in Ni-TCNQ and antiferromagnetism in Mn-TCNQ. 

Being aware that our DFT calculations underestimate the HOMO-LUMO gap of the TCNQ molecules, \cite{baldea} it is worth to mention that our estimated values for the exchange coupling constants (J) below are only an order of magnitude estimate. This is due to the approximation of considering Kohn-Sham (K-S) eigenvalues as true eigenvalues with physical meaning. Strictly speaking, only the last occupied  K-S orbital has physical meaning, which in our systems is the minority LUMO that is hybridized to a minor or greater extent with 3d atomic orbitals of the Mn or Ni transition metal atoms, respectively. In practice, this approximation affects more the value of the hoppings (t) than the energy denominators in our 2nd and 4th order perturbative models described in the next sections to estimate J for Ni-TCNQ and Mn-TCNQ. Therefore, we insist in the limited validity of our accuracy in determining the values of J, the important point being that that they differ by two orders of magnitude and in their sign that corresponds to FM coupling in Ni-TCNQ and very weak AFM coupling in Mn-TCNQ.

\bigskip

\section{\protect\bigskip Model for Ni-TCNQ ferromagnetism}

The mechanism of ferromagnetism in Ni-TCNQ is similar to the one described by Zener in 1951.~\cite{Zener} 
Localized spins and itinerant spin density are coupled via the Heisenberg exchange interaction,
which assumes the ferromagnetic sign if the hybridization of the conduction electrons (dispersive LUMO band) 
with a doubly-occupied or empty $d$ orbital of the magnetic center ($3d_{xz}$~and~$3d_{yz}$) is sufficiently strong.
Indeed, owing to Hund's rule in the $d$ shell,
it is energetically favorable to induce a spin polarization parallel to the $d$-shell spin.
The itinerant spin density, however, forms at an energy penalty determined by the dispersion of the conduction band;
the larger the density of states at the Fermi level,
the easier is for the itinerant spin density to form.

From the DFT results,
we learn that each Ni atom in the Ni-TCNQ network hosts a local spin $S=1/2$, localized in its $d_{xy}$ orbital, 
whereas the LUMOs of the TCNQ molecules couple together to form a band of itinerant electrons.
To describe the magnetic properties of the Ni-TCNQ network,
we employ the model Hamiltonian
\begin{equation}
H=-J\sum_{\left\langle i j \right\rangle}\bm{S}_{i}\cdot\bm{s}(\bm{r}_{j})
+\sum_{\bm{k}\sigma}\varepsilon_{\bm{k}}c^\dagger_{\bm{k}\sigma}c_{\bm{k}\sigma}+
H_Z,
\label{eqHamNi}
\end{equation}
where $J$ is the exchange coupling constant between the Ni spin $\bm{S}_{i}$
and the itinerant spin density $\bm{s}(\bm{r}_{j})$ at the TCNQ site $\bm{r}_{j}$.
For each Ni site $i$, the sum over $j$ runs over its $4$ neighbouring TCNQ molecules.
The spin density operator reads
\begin{equation}
\bm{s}(\bm{r}_{\bm{j}})=\frac{1}{2N}\sum_{\bm{k}\bm{k}'\sigma\sigma'}
\bm{\tau}_{\sigma\sigma'}
e^{i(\bm{k}'-\bm{k})\cdot\bm{r}_j}
c^\dagger_{\bm{k}\sigma}c_{\bm{k}'\sigma'},
\end{equation}
where $c^\dagger_{\bm{k}\sigma}$ creates an electron with wave vector $\bm{k}=(k_x,k_y)$ and spin 
$\sigma=\uparrow,\downarrow$ in the conduction band, $N$ is the number of lattice sites,
and $\bm{\tau}=(\tau_x,\tau_y,\tau_z)$ is a set of Pauli matrices.
The conduction band has dispersion
\begin{equation}
\varepsilon(k_x,k_y)=-2t_x\cos(k_xa_x)-2t_y\cos(k_ya_y)-4t'\cos(k_xa_x)\cos(k_ya_y),
\label{eqspectrum}
\end{equation}
where $t_x$ and $t_y$ are the tunneling amplitudes between LUMOs of neighbouring molecules along $x$ and along $y$, respectively.
The last term in Eq.~(\ref{eqspectrum}) arises due to the next-to-nearest-neighbour coupling, 
such as the coupling mediated by the $d_{xz}$ and $d_{yz}$ orbitals of the Ni atoms (see further).
We emphasize that, due to symmetry constraints, among the Ni $d$ orbitals, only $d_{xz}$ and $d_{yz}$ hybridize appreciably 
to the TCNQ LUMO (essentially, atomic $p_{z}$ orbitals) and,
therefore, play an important role in determining the strength of magnetism.
The last term in Eq.~(\ref{eqHamNi}) stands for the Zeeman interaction, for which we take
$H_Z=g \mu_B\sum_i\bm{S}_i\cdot\bm{B}+g \mu_B\sum_j\bm{s}(\bm{r}_j)\cdot\bm{B}$,
with the g factor $g=2$ and the magnetic field $\bm{B}=(0,0,-B)$.

\begin{figure}[ht]
\includegraphics[width=.45\textwidth]{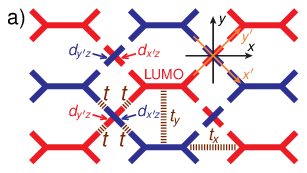}
\hfill
\includegraphics[width=.45\textwidth]{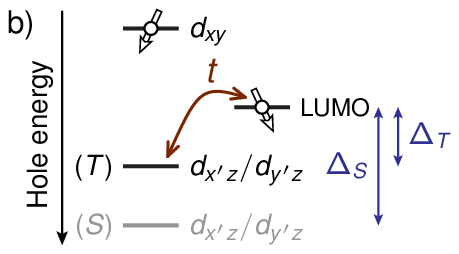}\\
\caption{\label{FigFMNi}
(color online) Mechanism of ferromagnetic interaction in Ni-TCNQ.
(a) Schematic top view of the coordination network, 
showing tunnel coupling between relevant orbitals.
Each Ni atom is represented by its $d_{x'z}$ and $d_{y'z}$ orbitals
chosen in the coordinate frame $(x',y')$.
Each $d_{x'z}$ (or $d_{y'z}$) orbital couples, with the tunneling amplitude $t$,  
to its two neighbouring LUMOs on one of the two sublattices distinguished by blue and red colors.
The tunneling amplitudes $t_x$ and $t_y$ give the 
coupling between the two intercalated sublattices.
(b) Energy diagram illustrating the origin of the exchange coupling in the hole representation.
The local spin is due to a hole residing in the $d_{xy}$ orbital.
The LUMO hole hybridizes with the $d_{x'z}$ (or $d_{y'z}$) orbital
due to the tunnel coupling with the amplitude $t$.
Owing to Hund's rule, 
$d_{x'z}/d_{y'z}$ is closer in energy to the LUMO
when the two holes form a triplet state (position $T$)
than when they form a singlet (position $S$).
The difference in energies gained by hybridization
in the triplet and singlet sectors gives the exchange constant $J$.
}
\end{figure}

To keep our discussion simple, we dispense with the splitting between the
$d_{xz}$ and $d_{yz}$ orbitals induced by the ligand field~\cite{note1}.
We thus adopt $\pi/4$-rotated orbitals,
$d_{x'z}=(d_{xz}-d_{yz})/\sqrt{2}$ 
and $d_{y'z}=(d_{xz}+d_{yz})/\sqrt{2}$, and show
the origin of the coupling constants $J$ and $t'$ in Figs.~\ref{FigFMNi} (a) and (b).
In Fig.~\ref{FigFMNi}~(a),
we represent schematically each magnetic center by its $d_{x'z}$ and $d_{y'z}$ orbitals
and each TCNQ molecule by its LUMO.
Neighbouring molecule LUMOs are tunnel coupled both directly, with the tunneling amplitudes $t_x$ and $t_y$,
and indirectly, via the magnetic center.
In the latter case, the tunneling amplitude between $d_{x'z}$ (or $d_{y'z}$) and LUMO is denoted by $t$.
The simplest situation arises when the direct coupling is absent ($t_x=t_y=0$)
and the itinerant electrons fall into two independent Fermi seas,
formed by two intercalated sublattices, as differentiated by the blue and red colors in Fig.~\ref{FigFMNi}~(a).
The two Fermi seas interact with 
the lattice of local spins, hosted by the $d_{xy}$ orbitals of the Ni atoms, not shown.
In Fig.~\ref{FigFMNi}~(b), we show the origin of this exchange interaction,
using the language of holes.
The coupling constant $J$ arises from virtual hops of the LUMO hole onto the $d_{x'z}$ (or $d_{y'z}$) orbital.
Because of Hund's rule, the energy denominator for the virtual transition depends on
whether a triplet $(T)$ or a singlet $(S)$ is formed on the magnetic center.
By perturbation theory, the exchange constant reads
$J=t^2\left(1/\Delta_T-1/\Delta_S\right)$, where
$\Delta_T$ and $\Delta_S$ are the energies depicted in Fig.~\ref{FigFMNi}~(b).
Similarly, the tunneling across the magnetic center,
mediated by the $d_{x'z}$ (or $d_{y'z}$) orbital,
has amplitude $t'=-t^2\left(3/4\Delta_T+1/4\Delta_S\right)$, 
where the minus sign signifies an anti-bonding coupling.
In addition to the exchange coupling and the mediated tunneling,
other terms arise in perturbation theory, but are not present in Eq.~(\ref{eqHamNi}).
Although those terms~\cite{note2} may account for some finer features seen in the DFT results,
such as the spin dependence of the width of the LUMO band,
they are generally unimportant for explaining the experiment.

In the Methods section, we describe two different methods for extracting the value of J for this model of ferromagnetism, one uses parameters extracted from the DFT calculations and the other is based on the fitting of measured magnetization curves using the Weiss theory. Both methods yield different J values but they are of the same order of magnitude. However, J values extracted from Monte Carlo simulations assuming an ensemble of localized spins are typically an order of magnitude smaller \cite{nasiba2013} and, thus, reflect that the physical meaning of J is different in our model with itinerant spin density.
The value of $J$ extracted from the DFT calculation ($J=22\,\textrm{meV}$) 
is several times larger than the one obtained from 
fitting the magnetization curve with the help of the Weiss theory ($J=6-11\,\textrm{meV}$).
While there are many possible reasons for this discrepancy,
we would like to emphasize that the Weiss theory tends to exaggerate the strength of ferromagnetic effects,
since it does not account for the possibility of exciting spin waves \cite{yosida}.
Indeed, the spin flip-flop terms in Eq.~(\ref{eqHamNi}), $-J\left[S_i^xs^x(\bm{r}_j)+S_i^ys^y(\bm{r}_j)\right]$, 
are disregarded in the Weiss theory, making, thus, effectively no distinction between the Heisenberg and  
Ising types of spin-spin interaction.
In 2D, the presence/absence of the flip-flop terms makes a qualitative difference at low temperatures, resulting in
absence/presence of magnetic order.
As a result, $T_c=0$ for the model in Eq.~(\ref{eqHamNi}), whereas $T_c>0$ for its Ising-type version, 
in which the flip-flop terms are absent.

Furthermore, we remark that the flip-flop terms are accounted for in the spin-wave theory. 
In 2D, however, the spin-wave expansion works only in the presence 
of a sufficiently strong magnetic field and at low temperatures,
such that the average spin $S_z$ is close to $1/2$.
In this region of $B$, 
the magnetization curve is nearly flat and
the accuracy of such a fitting (by spin-wave theory) is poor.
Note that the experimental data, i.e., the XMCD intensities, are only proportional to the magnetization;
the fitting procedure uses, thus, an arbitrary scaling factor to rescale the measured curve as desired.

One might envision that the magnetization curve calculated within
a more accurate theory agrees well with the one obtained using the Weiss theory,
if $J$ is replaced in the latter by a running coupling constant $J(T)$.
Then, this effective coupling $J(T)$ should tend to $J$ at high temperatures
and to zero at low temperatures. While this is only a conjeccture, we remark that such a running coupling constant 
readily occurs in this model due to the build up of Kondo correlations. Since $J$ is ferromagnetic, the scaling due 
to the Kondo correlations acts to reduce the magnitude of $J$~\cite{Hewson}. 
However, this reduction is rather weak (a factor of 2 at most) and cannot validate the use of the Weiss theory at arbitrary low temperature. 
Nevertheless, the agreement between the Weiss theory and the measured data is very good at T=8 K (see Fig.~\ref{FigMagFit} )

\bigskip

\section{\protect\bigskip Model for Mn-TCNQ antiferromagnetism}

\begin{figure}[ht]
\includegraphics[width=.45\textwidth]{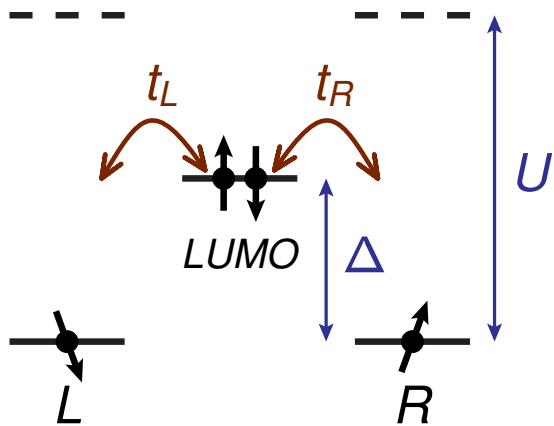}
\caption{\label{FigAF}
(color online) 
Energy diagram illustrating the origin of the AFM exchange coupling.
}
\end{figure}

The mechanism of anti-ferromagnetism in Mn-TCNQ is similar to the one described by Anderson in 1950.~\cite{Anderson}
Localized spins in Mn $d$ shells 
interact between one another via a superexchange mechanism, 
in which a $d$-shell electron (or hole) of a Mn atom tunnels in a virtual transition 
onto the ligand, whereon it experiences the correlation energy with the $d$-shell of another Mn atom adjacent to the ligand. 
In order to explain the basic mechanism that we take into account, we simplify the problem by retaining only one
orbital per Mn atom, considering, thus, the case of $S=1/2$ at each magnetic center. 
As for the ligand, we retain only its LUMO. 
The energy diagram for the interaction of two localized spins via the LUMO of the ligand 
is shown in Fig.~\ref{FigAF}.
Since the LUMO is doubly occupied with electrons, the superexchange occurs as a result of virtual transitions of the LUMO electrons onto the $d$-shell orbitals.
The coupling between the two localized spins at Mn atom sites has the form of the Heisenberg exchange interaction
\begin{equation}
H=-J \bm{S}_L\cdot\bm{S}_R,
\end{equation}
where $J$ is the coupling constant obtained from superexchange. 

In order to estimate $J$, we assume that the Coulomb interaction between electrons is local, 
{\em i.e.} electrons interact via an onsite Coulomb repulsion, such as in the Hubbard model. 
This assumption is motivated by the fact that the ligand is a relatively large molecule, 
for which the principal source of exchange comes from tunneling rather than Coulomb exchange matrix elements.
Indeed, the matrix elements of the Coulomb exchange taken between the Mn $d$-shell and 
the LUMO decrease with the size of the ligand.
Furthermore, the presence of the underlying substrate effectively screens the Coulomb interaction, making it local. 
Thus, we estimate $J$ to be

\begin{equation}
J=-\frac{4t^4}{(U-\Delta)^2}\left(\frac{1}{U-\Delta}+\frac{1}{U}\right),
\label{eqJN2DU}
\end{equation}
where $\Delta$ is the energy distance shown in Fig.~\ref{FigAF} and $U$ is the Coulomb repulsion on the site of the $d$ orbital.
To generalize Eq.~(\ref{eqJN2DU}) to the case of Mn-TCNQ, we need to introduce a factor $1/(2S)^2$ on the right-hand side, where
$S=5/2$ is the spin of the Mn atom.
We remark that the tunnel coupling between the Mn $d$-shell and the TCNQ-LUMO takes place only via
one of the $d_{zx'}$ or $d_{xy'}$ orbitals, as illustrated in the diagram in Fig.~\ref{FigFMNi}a; the diagram applies also for the Mn case.
Additionally, we remark that the superexchange between two neighbouring Mn spins on the lattice differs from the one
illustrated in Fig.~\ref{FigAF} by the possibility of involving two (and not one) LUMO orbitals.
Thus, superexchange via the red and blue sublattices in Fig.~\ref{FigFMNi}a are both possible.
However, this difference amounts only to a factor of $2$ in the end result, since the two paths do not interfere.
By analysing the DFT data, we deduce $\Delta\approx 4.0\,\textrm{eV}$ and $U=7.5\,\textrm{eV}$
and estimate $J=0.04\,\textrm{meV}$ for the nearest neighbours and $J'=0.02\,\textrm{meV}$ for the next-to-nearest neighbours.

\bigskip

\section{\protect\bigskip Conclusion}

In conclusion, our XMCD data for Ni-TCNQ and Mn-TCNQ networks on Au(111) with the same 1:1 stoichiometry and 4-fold coordination show very distinct magnetic behavior: only the Ni-TCNQ network shows ferromagnetic coupling between the Ni spin magnetic moments.  

With the help of first-principles DFT+U calculations we have been able to explain the qualitative differences between the two systems and extract parameters for the perturbative model Hamiltonians. These permit an order of magnitude estimate of the exchange coupling constants (J), no matter whether DFT+U calculations have limitations due to the underestimation of the HOMO-LUMO gap and the choice of the U parameter value. 

A fit of the measured magnetization curve for Ni-TCNQ assuming S=1 magnetic moments localized at Ni sites \cite{nasiba2013}, or S=1/2 magnetic moments at the Ni and TCNQ sites, that are coupled through a Heisenberg exchange gives J values which are an order of magnitude smaller than our J estimates, reflecting that the physical meaning of J is different in our model with itinerant spin density. However, in the Mn-TCNQ case the assumption of S=5/2 spin magnetic moments localized at the Mn sites seems to be well justified and, therefore, also the meaning of the corresponding value of J.

More importantly, we have found that the reason for the appearance of ferromagnetism in Ni-TCNQ is the existence of Heisenberg exchange coupling between spins localized at Ni sites and the itinerant spin density that appears due to the spin polarization of the LUMO band, hybridized with Ni(3d) states close to the Fermi level. Additionally, we have found that in Mn-TCNQ, the spin magnetic moments are localized at the Mn sites and, furthermore, they are very weakly anti-ferromagnetically coupled, in agreement with the observed behavior (essentially, paramagnetic at 8 K). 

These two cases can be considered as two opposite limiting cases showing FM and weak AFM coupling but, in principle, there would exist other situations that may give rise to different magnetic phases, e.g. ferrimagnetic coupling, in which spin magnetic moments at the metal atoms have different magnitude and direction than the spins of the organic ligands\cite{miller2012}. Further studies of this sort of systems, in which transition metal atoms form long range order two-dimensional networks with different size and shape organic ligands, would allow to explore the role of different coordination and stoichiometry. 

\bigskip

\section{\protect\bigskip Methods}

The STM experiments were carried out in an ultra-high vacuum chamber with a base pressure of better than $2\times 10^{-10}$ mbar in the preparation chamber and lower than $1\times 10 ^{-11}$ mbar in the STM. The Au(111) surface was cleaned by repeated cycles of Ar$^+$ sputtering and subsequent annealing to 800 K. TCNQ (98 $\%$ purity, Aldrich) was deposited by organic molecular-beam epitaxy (OMBE) from a resistively heated quartz crucible at a sublimation temperature of 408 K onto the clean Au(111) surface kept at room temperature. The coverage of molecules was controlled to be below one monolayer. Ni and Mn were subsequently deposited by an electron-beam heating evaporator at a flux of $ \sim$ 0.01 ML/min on top of the TCNQ adlayer held at 350 K to promote the network formation. The substrate was subsequently transferred to the low-temperature STM and cooled to 5 K. STM images were acquired with typical parameters of I = 0.1-1 nA and U= $\pm$ 0.5-1.2 V.
Polarization-dependent XAS experiments were performed at the beamline ID08 of the European Synchrotron Radiation Facility using total electron yield detection. Magnetic fields were applied collinear with the photon beam at sample temperatures between 8 and 300 K. A linear background was subtracted for clarity. The metal substrates were prepared by sputter-anneal cycles. The preparation of the metal-organic networks followed the protocols established in the STM lab. The sample preparation was verified by STM before transferring the samples to the XMCD chamber without breaking the vacuum. 

Calculations for Ni-TCNQ and Mn-TCNQ were performed with the Vienna Ab Initio Simulation Package (VASP) \cite{vasp1,vasp2}. These systems were modelled with a periodic supercell, the ion-electron interaction was described with the Projector Augmented-Wave (PAW) method \cite{paw}, whereas the exchange and correlation potential was taken into account by the Generalized Gradient Approximation(GGA)\cite{gga}. In both systems the plane wave expansion considers a kinetic energy cut-off of 280 eV. To satisfy the summations in the reciprocal space for the Brillouin zone a mesh of $4\times 6$ k points in the $1\times 1$ unit cell was chosen. Two planar (XY-plane) geometries were considered for each system, (a) the rectangular $1\times 1$ cell, from where the PDOS was extracted and (b) the checker-board geometry in a $2\times 1$ cell that allowed to estimate the FM or AFM coupling on each system. Ni-TCNQ and Mn-TCNQ networks were optimized both in lattice constants and atomic positions, assuming a convergence criterion of 0.01 eV/\AA\ in the rectangular $1\times 1$ cell and 0.05 eV/\AA\ in the $2\times 1$ cell with checker-board geometry. For all calculations the electronic convergence criterion was $1\times 10^{-6}$ eV. With the aim to describe properly the d electrons in Ni and Mn metal centres,  spin polarized calculations in the DFT + U approach \cite{dudarev} with a value of U=4 eV were performed. 
We have checked that varying the value of U in the range 3 to 5 eV does not change the values of the Ni and Mn magnetic moments appreciably neither the corresponding 3d level occupations, in particular that of the Ni($3d_{xz}$) orbital that crosses the Fermi level. Therefore, our conclusions do not depend on the choice of the particular value of U in this range.

\subsection{Extraction of J fitting magnetization curves}

\begin{figure}[ht]
\includegraphics[width=.45\textwidth]{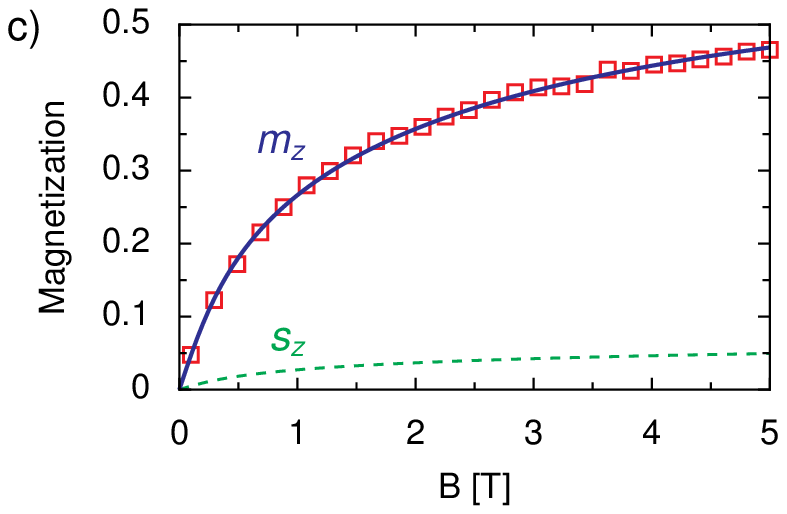}
\caption{\label{FigMagFit}
The data set of Fig.~\ref{FIG1}~(e) (squares) fitted
with the help of the Weiss theory in Eq.~(\ref{eqWeiss}).
The total magnetization $m_z=S_z+s_z$ (solid line)
as well as its itinerant component $s_z$ (dashed line)
are plotted versus $B$ for the parameter values
$J=5.55\,\textrm{meV}$ and $W=100\,\textrm{meV}$.
}
\end{figure}

We consider the Weiss theory~\cite{AshcroftMermin} for the model in Eq.~(\ref{eqHamNi}).
Under the assumption that the magnetization is homogeneous,
the average magnetic moment per unit cell is
$m_z=S_z+s_z$,
where $S_z\equiv\left\langle S_i^z\right\rangle$ and  $s_z\equiv\left\langle s_z(\bm{r}_j)\right\rangle$
are found by solving two coupled equations,
\begin{eqnarray}
&&S_z=\frac{1}{2}\tanh\left(\frac{\epsilon_Z+2Js_z}{T}\right),\nonumber\\
&&s_z=\frac{1}{2}\int d\varepsilon 
f(\varepsilon-\mu)\left[
\nu_\uparrow(\varepsilon)-\nu_\downarrow(\varepsilon)
\right]\approx \frac{\epsilon_Z+2JS_z}{W}.
\label{eqWeiss}
\end{eqnarray}
Here, $\epsilon_Z=\frac{1}{2}g\mu_BB$ is the Zeeman energy,
$f(\varepsilon)$ is the Fermi-Dirac distribution function, and
$\nu_{\uparrow/\downarrow}(\varepsilon)=\nu(\varepsilon\pm\epsilon_Z\pm2JS_z)$,
with $\nu(\varepsilon)$ being the density of states of the itinerant carriers.
For simplicity, we approximate the integral in Eq.~(\ref{eqWeiss}) by the mean-value theorem, assuming that
$\nu(\varepsilon)$  changes weakly on the scale of $\epsilon_Z+2JS_z$.
The resulting effective band width is then approximated by the density of states at the Fermi level,
$W\approx 1/\nu(\mu)$,
and the chemical potential $\mu$ is assumed to be independent of $B$\cite{note3}.  
With the help of this simple theory, which has $J$ and $W$ as unknown parameters,
we obtain magnetization curves similar to those measured for Ni-TCNQ.
An example is shown in Fig.~\ref{FigMagFit},
where, for $J= 5.55\,\textrm{meV}$ and $W=100\,\textrm{meV}$, we
reproduce the shape of the XMCD curve measured for normal x-ray incidence (same data set as in Fig.~\ref{FIG1}~(e)).
The XMCD signal is multiplied by a constant factor, which is regarded as a fitting parameter.
Furthermore, similar fits to the same data set can be obtained for different combinations of values of $J$ and $W$.
For instance, we swept $W$ from $20\,\textrm{meV}$ to $500\,\textrm{meV}$ and 
for each value of $W$ we could find a value of $J$ for which a
fit as good as the one in Fig.~\ref{FigFMNi}~© was produced.
The value of $J$ extracted from the fitting procedure scales as $J\propto\sqrt{W}$.
On the other hand, one finds from Eq.~(\ref{eqWeiss}) that the critical temperature in the Weiss
theory is $T_c=2J^2/W$.
Thus, the best-fit procedure allows us to determine only $T_c$ 
rather than $J$ and $W$ separately.
We find that the extracted value of $T_c$ depends weakly on $W$, 
varying from $0.61\,\textrm{meV}$ to $0.62\,\textrm{meV}$ during the sweep.
It should be noted, however, that 
the Weiss theory is at verge of its applicability,
since the temperature in the experiment is close to
the extracted value for the critical temperature, $T_c\approx 7\,\textrm{K}$.
For lower temperatures, $0<T<T_c$,
the Weiss theory predicts a non-zero average magnetization at $B=0$,
which is incorrect for the model in Eq.~(\ref{eqHamNi}).
A more accurate theory lowers this critical temperature
down to $T_c=0$.
Nevertheless, the Weiss theory produces a scale for the bending of the magnetization curve,
$\epsilon_Z\sim (T-T_c)/(1+2J/W)$,
that is lower than the scale at which the spin-$1/2$ Brillouin function bends, 
$\epsilon_Z\sim T$.

\subsection{Extraction of J from DFT calculations}

In order to give an independent estimate for $J$ and $W$, we analyze 
the results of the DFT calculations performed for the Ni-TCNQ network.
We find that already the simplest DFT calculation, in which the Brillouin 
zone is spanned by a single $k$-point ($\Gamma$-only calculation), 
suffices to estimate the values of $t$, $\Delta_S$, and $\Delta_T$.
From the level positions and the hybridization strength of the
LUMO with the $d_{xz}$ and $d_{yz}$ orbitals, we deduce
$t\approx 0.2\,\textrm{eV}$, $\Delta_S\gtrsim 2.8\,\textrm{eV}$, and $\Delta_T\approx 1.1\,\textrm{eV}$.
It should be noted here that
a $1$-$k$ point DFT calculation features an enhanced hybridization strength for some of the orbitals 
as compared to a multi-$k$ point calculation.
We have accounted for this enhancement by dividing the tunneling amplitude between the LUMO and the $d_{xz}$ orbital by $2$;
the $d_{yz}$ orbital does not couple to the LUMO in the $\Gamma$-only calculation.
This doubling of tunnel amplitude has its origin in the fact that the $d_{xz}$ orbital couples at its both ends to one and the same LUMO,
resulting in an enhanced coherence, {\em i.e.} constructive interference.
The fact that the $d_{yz}$ orbital decouples can be attributed in a similar way to destructive interference.
Furthermore, the transition from $\{d_{xz},d_{yz}\}$ to $\{d_{x'z},d_{y'z}\}$ introduces an additional factor of $1/\sqrt{2}$.
Thus, the estimate for $t$ was obtained by dividing the tunnel amplitude between the LUMO and $d_{xz}$
by $2\sqrt{2}$.

We performed also multi-$k$ point DFT calculations, although they are, {\em per se}, more difficult to analyze.
We note only that, if one averages the projected DOS over the Brillouin zone in a multi-$k$ point calculation,
then the interference terms cancel out up to terms of order $1/N_k$, 
where $N_k$ is the number of $k$ points used in the DFT calculation.
Thus, for $N_k\gg 1$, the coupling of the $d_{xz}$ orbital to its $4$ nearest-neighbor LUMOs
can be added incoherently, yielding an admixture strength of $4(t/\sqrt{2})^2/\Delta^2$, where
$\Delta$ is the energy distance between the LUMO and the $d_{xz}$ orbital.
This is to be contrasted with the $1$-$k$ point case discussed above, 
for which one has an admixture strength of $(4t/\sqrt{2})^2/\Delta^2$ arising from coherent addition.
In practice, we performed a $24$-$k$ point DFT calculation and found that the values of $t$ extracted by both methods 
coincide within expected accuracy.

Having extracted $t$, $\Delta_S$, and $\Delta_T$ from the projected DOS, 
we estimate
$J\approx 22\,\textrm{meV}$ and $t'\approx -31\,\textrm{meV}$ using the expressions for 
$J$ and $t'$ :
$J=t^2\left(1/\Delta_T-1/\Delta_S\right)$ and $t'=-t^2\left(3/4\Delta_T+1/4\Delta_S\right)$.
To determine the remaining unknown parameters, $t_x$ and $t_y$,
we compare the spectrum of the majority LUMO band computed in DFT
and the expression in Eq.~(\ref{eqspectrum}).
The two spectra agree well for 
$t_x\approx-32\,\textrm{meV}$,
$t_y\approx42\,\textrm{meV}$, and
$t'\approx-26\,\textrm{meV}$.
Note that the difference between the two values estimated for $t'$ is
about $J/4$ and may be attributed to the fact that we dispensed with some terms~\cite{note2}
when deriving Eq.~(\ref{eqHamNi}).
A more rigorous calculation shows that
the spectrum of the majority LUMO band is given by the expression in Eq.~(\ref{eqspectrum})
with $t'\to t'_\uparrow=-t^2\left(1/2\Delta_T+1/2\Delta_S\right)\approx-25\,\textrm{meV}$.
Similarly, for the minority LUMO band, one expects 
$t'\to t'_\downarrow=-t^2/\Delta_T\approx-36\,\textrm{meV}$, 
{\em i.e.} the minority LUMO band is somewhat wider than its majority counterpart.
However, the DFT calculation shows also that the minority LUMO band mixes strongly with the $d_{xz}$ orbital,
since the $d_{xz}$ orbital lies close in energy to the LUMO.
Therefore, our results derived with the help of perturbation theory are only qualitatively correct in this case.
Nevertheless, a rough estimate for $W$ can be given either from the projected DOS 
or from the DOS evaluated for the dispersion relation in Eq.~(\ref{eqspectrum}).
The latter method yields $W\approx 113\,\textrm{meV}$, whereas the former $W\lesssim 400\,\textrm{meV}$.

\bigskip

\section{\protect\bigskip Acknowledgements}

M. N. F. and A. A. thank MINECO (grant number FIS2010-19609-C02-01)  and  Eusko Jaurlaritza - UPV/EHU (grant number IT-756-13) for financial support  and DIPC for providing us with computational resources of its Computer Center. V. N. G. was supported by the Spanish Ministry of Economy and Competitiveness under Project No. FIS2011-28851-C02-02. We thank the ESRF for the provision of beam time to do the X-ray absorption experiments.



\begin{thebibliography}{99}
\bibitem{note1}
Indeed, this splitting plays a minor role in our theory
and can be taken into account at the very end 
by substituting $2\epsilon_{d_{xz}}\epsilon_{d_{yz}}/(\epsilon_{d_{xz}}+\epsilon_{d_{yz}})$
for the effective position of the $\{d_{xz},d_{yz}\}$ multiplet.
Here, $\epsilon_{d_{xz}}$ and $\epsilon_{d_{yz}}$ are 
the energies of the $d_{xz}$ and $d_{yz}$ orbitals, respectively.
\bibitem{note2}
These include a term renormalising the LUMO energy ($\varepsilon\to\varepsilon-4t'$)
and a term that can be combined with $t'$ to produce a
spin-dependent tunneling with the amplitude
$t'_{\sigma\sigma'}(\bm{r}_i)=t'\delta_{\sigma\sigma'}+J\bm{S}_i\cdot\bm{\tau}_{\sigma\sigma'}/2$,
where $\bm{r}_i$ is the position of the Ni atom mediating the tunneling.
In DFT, one fixes $\bm{S}_i\to \left\langle\bm{S}_i\right\rangle=(0,0,1/2)$, which leads to two tunneling amplitudes,
$t'_\uparrow=t'+J/4$ and $t'_\downarrow=t'-J/4$, respectively, for the majority and minority spins.
\bibitem{note3}
Strictly speaking, $\mu$ should be determined self-consistently from an electro-chemical equation,
taking into account that the charge 
$q\propto \int d\varepsilon f(\varepsilon-\mu)\left[
\nu_\uparrow(\varepsilon)+\nu_\downarrow(\varepsilon)\right]$
may vary with $\epsilon_Z+2JS_z$ when $\nu(\varepsilon)$ is a generic function.
Thus, $\mu$ may also vary with $\epsilon_Z+2JS_z$.
However, $\mu$ stays constant if $\nu(\varepsilon)$ is constant or has particle-hole symmetry around the Fermi level
within a range of energies $\left|\varepsilon-\mu\right|\leq\epsilon_Z+2JS_z$.
Moreover, 
since the metal-organic layer lies on a metal substrate,
the change of $\mu$ can be estimated as $\delta \mu=(e/C)\delta q$,
where $\delta q$ is the induced charge in the layer and 
$C$ is the capacitance of the parallel-plate capacitor formed by the layer and the substrate.
Obviously, $\delta \mu$ is extremely small
even in the absence of particle-hole symmetry, 
because the metal-organic layer lies at an atomic distance from the substrate.
\end{thebibliography}
\end{document}